# Thermodynamic behavior of short oligonucleotides in microarray hybridizations can be described using Gibbs free energy in a nearest-neighbor model.

*Stefan Weckx[1,5], Enrico Carlon[2,3,4], Luc De Vuyst[5], Paul Van Hummelen[1,6,]**

[1] MicroArray Facility, VIB , Herestraat 49, B-3000 Leuven, Belgium

[2] Interdisciplinary Research Institute c/o IEMN, Cité Scientifique BP 60069, F-59652 Villeneuve d'Ascq, France

[3] Ecole Polytechnique Universitaire de Lille, Cité Scientifique, F-59655 Villeneuve d'Ascq, France

[4] Institute for Theoretical Physics, K.U.Leuven, Celestijnenlaan 200D, 3001 Leuven, Belgium.

[5] Research Group of Industrial Microbiology and Food Biotechnology (IMDO), Vrije Universiteit Brussel, Pleinlaan 2, B-1050 Brussels, Belgium

[6] MicroArray Facility, K.U.Leuven, Herestraat 49, B-3000 Leuven, Belgium

Corresponding author:

Paul Van Hummelen, PhD

MicroArray Facility (MAF)

VIB

Herestraat 49, Box 816

B-3000 Leuven, Belgium

Tel: +32 16 347939

Fax: +32 16 347940

E-mail: paul.vanhummelen@vib.be

**ABSTRACT**

While designing oligonucleotide-based microarrays, cross-hybridization between surface-bound oligos and non-intended labeled targets is probably the most difficult parameter to predict. Although literature describes rules-of-thumb concerning oligo length, overall similarity, and continuous stretches, the final behavior is difficult to predict. The aim of this study was to investigate the effect of well-defined mismatches on hybridization specificity using CodeLink Activated Slides, and to study quantitatively the relation between hybridization intensity and Gibbs free energy ($\Delta G$), taking the mismatches into account. Our data clearly showed a correlation between the hybridization intensity and $\Delta G$ of the oligos over three orders of magnitude for the hybridization intensity, which could be described by the Langmuir model. As $\Delta G$ was calculated according to the nearest-neighbor model, using values related to DNA hybridizations in solution, this study clearly shows that target-probe hybridizations on microarrays with a three-dimensional coating are in quantitative agreement with the corresponding reaction in solution. These results can be interesting for some practical applications. The correlation between intensity and $\Delta G$ can be used in quality control of microarray hybridizations by designing probes and corresponding RNA spikes with a range of $\Delta G$ values. Furthermore, this correlation might be of use to fine-tune oligonucleotide design algorithms in a way to improve the prediction of the influence of mismatching targets on microarray hybridizations.

# INTRODUCTION

In transcriptome research, microarrays are used to indicate genes expressed by a given cell type or organism. The overall expression pattern is in most cases compared to other patterns in order to reveal differentially expressed genes between different states or experimental conditions. Microarray technology is based on nucleotide-nucleotide hybridizations, where one molecule "the probe" is attached to the surface of a slide and the other molecule "the fluorescently labeled target" moves freely in solution and will bind to its complementary probe. To obtain a microarray with high sensitivity and specificity, the design of probe sequences is an important step. A good sensitivity results in intensities proportional to the real amount of labeled target in the hybridization solution. To obtain a good specificity, homology should be high for the intended target sequence, but low for all other target sequences in solution to avoid cross-hybridization of non-intended similar sequences.

Previous work described in literature present hybridization studies with respect to sensitivity, specificity, cross-hybridization, mismatches, oligo length, and number of oligos per gene [1-5]. Kane et al. [1] demonstrated that 50-mer oligonucleotides have a comparable sensitivity as PCR-derived probes of 300-400 nucleotides, and that 50-mers show good specificity as long as the similarity with non-intended sequences is < 75% or in the absence of a stretch of 15 perfectly matching bases. A comparable similarity was reported by Hughes et al. [2], who performed hybridizations with permutations-containing 60-mer oligonucleotides, and found that a similarity between probe and target of less than 77% resulted in a lack of signal. Although Tiquia et al. [4] showed a 10-fold higher hybridization intensity for 60-mers compared to 30-mers, according to the results of Relógio et al. [3], the length of oligonucleotides can be decreased to 30-mers with the preservation of a good balance between sensitivity and specificity. These results provide a set rules-of-thumb to be taken into account in the oligo design process: oligonucleotides should be at least 30 bp long, should have a maximal similarity of 75% with non-intended similar sequences and may not posses perfect matching stretches longer than 15 bp with these non-intended sequences.

Whereas the former results were based on the outcome of hybridization experiments only, thermodynamics has been taken into account in several other studies [6-19]. Indeed, hybridization of nucleotide molecules in solution is a well understood biophysical process driven by thermodynamics [20]. Experiments over the past two decades on oligonucleotide melting have provided precise measurements of thermodynamic parameters, such as the binding free energy (ΔG), involved in the process [21]. These experiments have validated a simple model known as the nearest-neighbor (NN) model, where the ΔG is calculated as the sum of ΔG values for each couple of two neighboring bases in a sequence [21]. The free energy for mismatching bases in DNA sequences has been investigated too, and ΔG values are described that can be used to calculate the overall ΔG of a mismatch-containing DNA sequence [22-23]. Although there is a general consensus of the validity of the nearest-neighbor model for perfect-matching duplexes, the case of mismatches has been more debated. Indeed, some studies reported that a single mismatch may have an effect that goes beyond the nearest-neighboring bases [24]. However, these effects have not yet been systematically quantified.

Although previous studies reported on the influence of ΔG on hybridization intensity [15,17,19], and on the use of ΔG to select suitable oligos [10,16], it has been subject to debate whether these thermodynamic parameters, determined for DNA sequences in solution, apply for microarray hybridizations where one of the molecules is attached to a surface, and hence prone to steric hindrance. Fotin *et al.* [6] performed one of the first studies to quantify molecular interaction in a gel pad microarray. They found a convincing linear relationship between microarray hybridization free energies (ΔG (chip)) and the corresponding free energies in solution (ΔG(solution)). Various thermodynamics-inspired approaches have also been invoked in the analysis of microarray data (see the review of Halperin *et al.* [25] and references therein) which focused in particular to Affymetrix GeneChips. These approaches rely on the so-called Langmuir isotherm, but there is no consensus whether one can correlate ΔG (chip) and ΔG(solution). Pozhitkov *et al.* [18] reported a poor statistical relationship between hybridization intensity and ΔG(solution), and also stated that microarray-based hybridizations using mismatch-containing oligos did differ from solution-based hybridizations.

These somewhat controversial results call for more experimental investigations aimed at a better understanding and at a quantification of molecular interactions in microarrays. The aim of the present study was to assess the effect of well-defined mismatches on hybridization specificity using spotted microarray consisting of eight oligo sets with varying GC content, and to study quantitatively the relation between hybridization intensity and ΔG(solution), calculated from the nearest-neighbor model, taking mismatches into account.

## MATERIALS AND METHODS

**Oligo sets**

The genomic sequence of *Lactobacillus plantarum* was used to search for eight 30-mer sequences with a GC content ranging from 10% to 80% (Table 1). From each oligo, eleven additional oligos were derived by introducing one, two or three mismatches at well-defined places, allowing evaluation of the mismatch position and inter-mismatch distance. The mismatches were of the type AA, CC, GG or TT. These twelve oligos made-up one oligo set. For each set, a 70-mer complementary target sequence was chosen, consisting of the 30-bp region of the 30-mer plus 20 bp up- and downstream of the 30-mer (Figure 1; Supporting Information). To study the influence of a spacer sequence between the oligo and the slide surface, a related oligo set was derived containing an additional 12-mer universal spacer at the 5' end. This universal spacer had no homology to any other sequence in these sets and was the same for all oligos. The 192 oligos were obtained from Eurogentec (Seraing, Belgium) with 5' amino modification. The eight 70-mer target sequences were obtained from the same company with a 5' Cy3 modification and were dissolved in TE buffer pH 8.0 till a final concentration of 100 μM.

**Microarray production**

The 192 oligos were diluted to a concentration of 25 pmol/μl in 100 mM Na-phosphate buffer, 0.000675% sarcosyl (Sigma-Aldrich, Milwaukee, Wisconsin, USA). A MicroGrid II spotter (Genomic Solutions, Huntingdon, UK) was used to spot the oligos using 10K pins on CodeLink Activated Slides (GE Healthcare, Bucks, United Kingdom). Each oligo was spotted 12 times on each slide using the same pin. After spotting, the slides were placed in a chamber containing saturated NaCl for min 24 h. The active groups on the slides were blocked with preheated blocking solution (50 mM ethanolamine, 0.1M Tris, pH = 9.0) at 55 °C for 30 min. Slides were rinsed twice with ultrapure water and washed with preheated 4xSSC-0.1% SDS at 55 °C for 30 min in a shaking water bath. Slides were rinsed with ultrapure water and dried by centrifugation at 800 rpm for 3 min.

**Hybridizations**

Hybridization mixtures containing one or several 70-mer targets were prepared in 210 μl hybridization buffer (GE Healthcare) containing 50% formamide (Sigma-Aldrich). The hybridization mixtures were denatured at 96 °C for 3 min and immediately put on ice for at least 5 min. Hybridization mixtures were kept at 42 °C for 5 min and subsequently centrifuged at 12,000 rpm for 5 min. Slides were hybridized at 45 °C during 12 h and washed in 1x SSC / 0.2% SDS for 10 min, in 0.1x SSC / 0.2% SDS for 10 min, and in 0.1x SSC 4 min, using the Lucidea Slidepro (LSP) hybridization station (GE Healthcare). Slides were dried by the LSP using isopropanol and air. Subsequently, slides were scanned using the Agilent scanner (Agilent Technologies, Palo Alto, California, USA) at 10 micron, and images were analyzed using ArrayVision v7 (GE Healthcare).

**Data processing**

Spot intensity was regarded as above background if the intensity was larger than the sum of the local background signals and twice the background's standard deviation. The average oligo intensity and average background were calculated when at least 10 of the 12 replicate spots had an intensity above background. The average value for the 12 replicate spots was calculated taking a confidence interval of 95% into account.

The Gibbs free energy was calculated from the nearest-neighbor model using the experimental stacking free energies obtained from DNA melting experiments in solution, as reported by Peyret *et al.* [22]. It is known [22] that terminal mismatches, e.g. mismatches close to double helix edges, are more stable than internal mismatches, although this effect has not yet been quantified. The values for oligos 10 and 12, containing a mismatch in the third nucleotide at the 3' and 5' ends (Figure 1), should therefore be seen as an approximation.

**The Langmuir model applied to microarray hybridizations**

The simplest thermodynamic model for hybridization in DNA microarrays is the Langmuir model [25], which predicts a spot intensity as given by Equation 1, where $I_0$ is the non-specific background signal, c the target concentration in solution, T the experimental temperature, R the universal gas constant and $-\Delta G$ the binding free energy. In the limit of small concentrations and weak free energies, $ce^{\Delta G/RT} \ll 1$, the denominator in equation 1 can be neglected. In this limit, the intensity is proportional to the target concentration (a measure of the gene expression level), which is usually assumed when analyzing microarray data. In the opposite limit, the spot saturates; there are no longer free single stranded probes available for hybridization and the intensity attains a maximal value $I = I_0 + A$. Usually one assumes that the hybridization is a two-state process, in which either target and probe are unbound, or fully bound so that the description of the process is restricted to two thermodynamical states.

# RESULTS

**Hybridizations and reproducibility**

In total, 18 hybridizations were performed using nine different target mixtures. One mixture contained all eight labeled target molecules simultaneously at a concentration of 3000 pM. The other mixtures contained targets 3, 4, 5, and 6 separately at a final concentration of 380 pM or 3.8 pM. Each of the mixtures was prepared and hybridized in duplicate.

The reproducibility of the hybridizations is illustrated in Figure 2. To calculate the correlation coefficient for the 3000 pM plot, spots with an intensity close to the maximal detectable intensity by the scanner were excluded. The hybridizations using all targets at a concentration of 3000 pM showed a correlation coefficient of 88%. The coefficients for the other hybridizations ranged from 85-99% (data not shown).

**GC content and spacer**

The bar charts in Figure 3 give a detailed view on the effect of the GC content showing log-transformed hybridization intensities (in Relative Fluorescent Units or RFU) from a 3000 pM hybridization for six representative types of oligos. Figure 3A shows the intensities for the perfect matching (PM) oligo (oligo 1) of all oligo sets. The intensities for all oligos except sets 1 and 2 were at the maximal detection limit of the scanner. Figure 3B shows the intensities for oligo 3 containing one mismatch (MM) in the middle of the oligo. The intensities for the low GC oligo sets 1 and 2 are low, only one out of four was above the background cut-off, the intensities for the oligo sets with a GC content of 50% (sets 3, 4, 5 and 6) were not at the maximal level, and the intensities of the high GC oligos were at the maximal detection level. Figures 3C, 3D and 3E show the intensities of three different types of oligos containing two MM. Oligo 10 (Figure 3C) had the two MM close to each other (3 bp in between) at the 3' end, the free end of the oligo. In oligo 6 (Figure 3D), the 2 MM were separated by a perfect matching stretch of 16 bp, and oligo 11 had the two MM close to each other (3 bp in between) near the middle of the oligo. All three bar charts show the same tendencies: the oligos with a low GC content (oligo sets 1 and 2) had low intensities, of which most were below the background cut-off. The oligos with a high GC content (oligo sets 7 and 8) had high intensities, nearly at the maximal level. The intensities of the oligos with a GC content of 50% showed a good variation in specificity. From these data, it became clear that oligo sets 3 and 5 showed overall higher intensities than oligo sets 4 and 6. Also, the intensities for oligo 10, containing a stretch of 23 perfect matching bases was higher than the intensities for oligo 6 containing a stretch of 16 perfect matching bases, which were in turn higher than the intensities for oligo 11, in which the 2 mismatches were close to each other in the middle of the oligo. Oligo 9 (Figure 3F), containing three MM, had intensities at the background level. Oligos with a high GC content were below the maximal level, but still considerably above the intensities of the other oligos.

All eight oligo sets consisted of a subset of 30-mer oligos and a subset of the 30-mer oligos with a 12-mer universal spacer at the 5' end. As becomes clear from the bar charts, the intensities of the 30-mers with universal spacer were slightly lower than the intensities of the 30-mer, although these differences were in most cases within the 95% confidence intervals.

**Relationship between binding free energy and hybridization intensity: the Langmuir model**

Figures 4-6 show log-transformed background-corrected hybridization intensities for the 30-mer oligonucleotides (in RFU, Y-axis) in function of the Gibbs free energy ($\Delta G$, X-axis), calculated using the nearest-neighbor model based on $\Delta G$ values for hybridization in solution. In what follows, if not explicitly stated, we use $\Delta G$ as shorthand for $\Delta G$(solution). The free energy is obtained from $\Delta H$ and $\Delta S$ parameters as described by Peyret *et al.* [22] using the formula $\Delta G = \Delta H - T \Delta S$ where $T = 318K$ (45 °C).

To evaluate the Langmuir model in a non-saturating environment, a line with slope $1/RT = 1.6$ mol/kcal (with $T = 318$ K and $R = 0.002$ kcal / (mol K)) was plotted as a guide to the eye.

Figure 4 contains two graphs, based on data of two different hybridizations at 3000 pM in which all eight targets were present. Only the data points for the oligo sets with a GC content of 50% were plotted (oligo sets 3-6). PM oligos (oligo 1 in Figure 1) were all at the maximal detection level. Also the data points of the one MM (1MM, oligos 3, 4, and 5 in Figure 1) oligos had high intensities, spread out over two log intervals, until saturation. The same spread is found for the two MM oligos with two MM close to each other, having a perfect matching stretch of 23 bp (2MM23, oligos 10 and 12 in Figure 1). The oligos with two MM and a perfect matching stretch of 15 bp (2MM15, oligos 6, 7, and 8 in Figure 1) did not reach the maximal level and were spread out over three log intervals. The oligos with two MM and a perfect matching stretch less than 15 bp (2MM-15, oligo 11 in Figure 1) were around $I=10^3$. The oligos with three MM (3MM, oligo 9 in Figure 1) had low intensities, some below the background cut-off (see also Figure 4). All oligos had a $\Delta G$ between 24 kcal/mol and 37.5 kcal/mol. Most of the data followed nicely the line with slope $1/RT$, representing a Langmuir isotherm. Only the data points for the PM oligos and some data points for the 1MM and 2MM23 oligos deviated from the line, because of signal saturation. In addition, in the middle of the intensity range, some data points for the 3MM and 2MM-15 oligos deviated from that line. Comparing both graphs reveals that the two distinct hybridizations had a high reproducibility.

Figure 5 contains data of four hybridizations using targets 3, 4, 5 and 6 individually at a concentration of 380 pM. Also here, the PM and 1MM oligos still had high intensities close to or at saturation. The data points of the 2MM23 oligos were spread over 3 log intervals ($10^2$ to $10^5$), where the highest values were similar to the intensities of the 1MM oligos. The 2MM15 and 2MM-15 oligos ranged over two log intervals, with the lower limit around or under the background cut-off (see also Figure 4). The 3MM oligos had lower intensities between $10^1$ and $10^2$. All data points representing an intensity higher than $10^2$ fitted the line with slope $1/RT$, representing the Langmuir isotherm. Also here, some outliers were observed for the 2MM-15 oligos.

Figure 6 contains data of the hybridizations with a target concentration of 3.8 pM. The four graphs represent data from separate hybridizations of the four targets 3, 4, 5 or 6. The overall intensities were much lower compared to Figures 5 and 6, with most of the data points having an intensity around background. Only for oligo sets 3 and 5, the data points for the 1MM and 2MM23 oligos were above background and were in reasonable agreement with the Langmuir isotherm.

**DISCUSSION**

When designing oligonucleotides to construct a microarray, high sensitivity and specificity of the oligonucleotides should be achieved. During the design process, cross-hybridization of non-intended target molecules should be taken into account, and in most cases be avoided. Therefore, it is important to be able to have an insight in the effect of mismatches on the overall hybridization, influencing both specificity and sensitivity. As DNA-DNA hybridization is a thermodynamically driven process, which is already studied in detail, models such as the nearest-neighbor (NN) model are extensively described [21]. The use of this model to approach the thermodynamics of microarray hybridizations is a subject of discussion, since one basic assumption is totally different: whereas the NN model is related to DNA molecules in solution [21], in the case of microarrays one of the nucleotide molecules is attached to a surface, so hybridization might be liable to steric hindrance or other factors, resulting in a behavior of the molecules that does not fit the NN model. However, several studies reported a correlation between the signal intensity and $\Delta G$ [15,17,19], but Pozhitkov et al. [18] claimed that this is not the case for mismatch-containing oligonucleotides, and that there is only a poor correlation between measured intensities and $\Delta G$.

To study the effect of well-defined mismatches on hybridization sensitivity and specificity, to evaluate whether the NN model is valid for microarray hybridizations, and to investigate a possible relationship between intensity and $\Delta G$, calculated using thermodynamic parameters determined for DNA-DNA hybridizations in solution, we designed eight oligo sets with varying GC content. Each set consisted of twelve 30-mers with and without a 12-mer spacer, containing zero, one, two, or three mismatches (Figure 1 and Supporting Information). Each of the 192 oligos were spotted twelve times on CodeLink Activated Slides, where each oligo was spotted by four different pins. The microarray was hybridized with different amounts of 70-mer labeled target sequences, simulating hybridizations with high and low expressed genes.

A first and obvious observation was that GC content was an important parameter in the sensitivity of oligonucleotides in general, and for mismatch-containing oligonucleotides in particular (Figure 3). At a low GC content of 10 to 33% (oligo sets 1 and 2), intensities were overall lower compared to the other oligo sets, even for the perfect matching oligo 1 (Figure 3, all bar charts). This means that oligonucleotides with this low GC content were not sensitive enough, resulting in suboptimal hybridization of the target. At a high GC content of 67 to 90% (oligo sets 7 and 8), almost all oligos showed intensities at the maximum detection level (Figure 3, all bar charts). The high GC content made the oligos less specific for mismatches; the strong hydrogen bonds between the high amount of cytosine and guanine overruled almost completely the unfavorable energy for mismatching bases. This makes oligos with a high GC content not specific, and will also result in a non-proper detection of the actual target amount.

A more balanced relation between sensitivity and specificity was observed for the oligo sets 3, 4, 5 and 6 with a GC content of 50%, where two effects became clear: the effect of the amount of mismatches and positions of those mismatches, and the effect of the ΔG free energy. For these oligo sets, we observed that one mismatch (Figure 3B) did not affect the intensity in such a way that it dropped under the background cut-off, at the hybridization and washing conditions that were used. This phenomenon has also been described for hybridizations using the Affymetrix platform [26]. For the oligos with two mismatches, the relative positions of these mismatches played a crucial role: the closer the mismatches and the longer the remaining perfect matching stretch of the oligo, the higher the intensity was (Figures 3C, 3D, and 3E). These results are in agreement with earlier studies [1,2] dealing with longer oligonucleotides. Oligo 9, containing three mismatches, showed in all oligo sets a clearly lower intensity, which suggests that, given this concentration, at least three mismatches were needed to avoid non-intended targets to hybridize under the hybridization and wash conditions used. Comparing the intensities for the four oligo sets 3, 4, 5 and 6 (Figures 3B, 3C, 3D, and 3E) it is notable that the intensities of oligo sets 3 and 5 were always higher compared to the intensities of oligo sets 4 and 6. This could be explained by the ΔG free energy, which was higher for oligo sets 3 and 5 (Table 1).

Data presented in Figures 4-6 clearly suggest a correlation between the hybridization intensity and ΔG, which can be described by the Langmuir model, and this for all three concentrations used. The ΔG was calculated using the nearest-neighbor model, taking specific values for mismatches into account. The resulting ΔG-values ranged between 24 kcal/mol and 37.5 kcal/mol. Consequently, we can state that, from a thermodynamics point of view, the hybridization intensity is solely dependent on ΔG, hybridization temperature, and concentration, as indicated by Equation 1 (the proportionality constant $A$ is unknown). Our results thus confirmed that the nearest-neighbor model, used to calculate ΔG (solution), is strongly correlated to the actual hybridization free energy ΔG(chip) where one of the molecules is attached to a surface.

In an early experiment on gel pad microarrays, Fotin *et al.* [6] indeed reported a linear correlation between the two Gibbs free energies: ΔG(chip) = a ΔG(solution) + b, where the parameters where found to be a = 1.1 ± 0.2 and b = -3.2 ± 0.4 kcal/mol, for hybridizations to 8-mer sequences. In our experiments an increase or decrease of the calculated values for ΔG by a global constant $b$, representing the potential differences between conditions in solution compared to conditions on a microarray, does not affect the Langmuir model since $b$ can be omitted by redefining the parameter $A$ in Equation 1 as $A' = A \exp(b/RT)$. Hence a value of the additive constant $b$ cannot be determined from the experimental data presented here, as the prefactor $A$ is unknown. There are several effects that may lead to an overall constant rescaling of free energies, for instance denaturants as formamide or DMSO in microarray hybridization buffers, are most likely to produce an overall shift of the free energies (as is the effect of monovalent salt on ΔG [21]). These effects would influence the parameter $b$ in the above formula.

The data presented in this paper allow a determination of the relative differences in ΔG's. The fact that the experimental data, when plotted in logarithmic scale, follow a slope that equals 1/RT, implies that the factor $a$ in the above formula equals 1. This is in agreement with the result of Fotin *et al.* [6] on gel pad microarrays for which a = 1.1 ± 0.2. This conclusion should however not be taken as a general statement valid for all types of microarrays. The slides used in these experiments (CodeLink Activated Slides) have a three-dimensional polymeric coating to which the oligo probes are attached. As a result, the probes are positioned further away from the surface, thereby giving the hybridization a solution-phase behavior, as described by Dorris *et al.* [8]. This three-dimensional coating also explains the fact that the 12-mer spacer did not increase the signal intensities, as also observed earlier by Fotin *et al.* [6], suggesting that all 30 bases were fully available for hybridization without the use of a spacer [8]. Other studies showing a beneficial effect of a spacer [27,28] used two-dimensional coatings. More in particular, recent analysis of Affymetrix GeneChips data [12,13,26] showed that the intensities vs. ΔG plots tend to follow straight lines as seen in Figures 4-6, although with a slope different from 1/RT, as expected from the experimental temperature value. This suggests a rescaling of ΔG on the GeneChips compared to the spotted microarrays due to steric hindrance, or other surface-induced effects, thereby suggesting that the factor *a* in the above formula relating ΔG(chip) to ΔG(solution) will be smaller than 1.

It is previously described that terminal mismatches are more stable than internal mismatches [23]. As could be observed from our data, the intensities of oligos 10 and 12 indeed systematically deviated from the expected thermodynamical behaviour. Although this effect has not been quantified yet, we found that by empirically subtracting a value of 2.5 kcal/mol from the total Gibbs free energy in order to compensate for this effect, the resulting hybridization free energy became close to the values that could be expected from the Langmuir model. It should be pointed out that the empirical correction is constant for all oligos 10 and 12, labeled as 2MM23 in figures 4, 5, and 6.

As mentioned in the introduction, the nearest-neighbor model might have some limitations when applied to oligos containing mismatches. Although Peyret *et al.* [22] provided data to calculate ΔG for mismatch-containing oligos, Hall *et al.* [24] reported evidence that the influence of the flanking sequence context goes beyond the nearest-neighbors. Given these findings, it is likely that two neighboring mismatches on the same sequence influence each other and result in a combined effect. As there is no precise quantification of these effects yet, the nearest-neighbor parameters of Peyret *et al.* [22] were used. The combined effects could be responsible for some of the data spread in figures 4, 5, and 6. However, we do not expect substantial aberrations due to these effects since in most of our oligos, mismatches are separated by stretches of at least eight bases (except probes 10, 11, and 12).

Finally, Figures 5 and 6 show tendencies similar to the ones in Figure 4 concerning the relation between the intensities and the presence of mismatches: the PM oligos had, as expected, the highest intensities, followed by the 1MM oligos and 2MM23 oligos. The other oligos with two mismatches, 2MM15 and 2MM-15, showed lower intensities that were above or under the background cut-off, depending on the concentration of target used. The oligos with three mismatches (3MM) showed low intensities for all concentrations. These data confirm that the rules-of-thumb from previous studies, concerning global identity and length of consecutive base stretches, also apply for short 30-mer oligonucleotides: a consecutive stretch of 23 bp gave overall higher intensities compared to oligos with a 15 bp stretch, that on their turn had overall higher intensities than oligos with 2MM without a 15 bp stretch.

Based on microarray hybridizations using mismatch-containing oligos, spotted on slides with three-dimensional surface coating, we showed that the fluorescence intensities fit the Langmuir model, and that this relationship is valid over three orders of magnitude for the hybridization intensity. These results, besides having their own importance, can be interesting for some practical applications. For instance, the correlation between intensity and ΔG can be used in quality control of microarray hybridizations. As described by van Bakel *et al.* [29], the best approach to determine the performance of microarray experiments, is by adding different, well-known amounts of external control RNA, or spikes, to the RNA of biological samples, where the microarray contains probes specific for these external RNA sequences. The current set of available spikes focuses on different concentration and ratios, mimicking a range of low to high differential expressed genes. However, a more stringent test of the hybridization performance could be done with spikes hybridizing oligo sets similar to those reported in this study, i.e. containing mismatches. The degree to which the Langmuir model is followed, could be a measure for the quality of the microarray hybridization.

Another important application could be for oligo design strategies. Current oligo design algorithms select a specific probe to be as different as possible from other homologous sequences. The main criteria are the overall homology and the presence of a perfect matching stretch of a sufficient length.

We believe that a more accurate criterion would be to maximize the difference in ΔG between the perfect matching target-probe complex and possible complexes consisting of the same probe but homologous targets, leading to one or more mismatches. As the mismatch free energy depends rather strongly on the identity of the mismatch and of the flanking nucleotides, it could happen that a target with two mismatches has a higher affinity to a given probe sequence compared to the target with a single mismatch. Hence, selecting probes only on the basis of the number of mismatches with potential cross-hybridizing target may not be the optimal criterion.

**ACKNOWLEDGEMENTS**

This work was financed by SBO project IWT-030263 of IWT-Vlaanderen. The authors are grateful to Joke Allemeersch for her contributions concerning the R environment, and Kizi Coeck, Ruth Maes and Tom Bogaert for their assistance with the microarray experiments.

**SUPPORTING INFORMATION AVAILABLE**

Sequence information of all eight oligo sets is available free of charge via the Internet at http://pubs.acs.org.

**Table 1:** Overview of GC content, Tm and binding free energy of the eight oligo sets used (determined for the perfect matching oligo). Tm is calculated a simple empirical formula (http://www.promega.com/biomath/), while ΔG is calculated from the nearest-neighbor model [22] at the hybridization temperature of 45 °C.

|  | perfect matching oligo (5' – 3') | GC (%) | Tm (°C) | $\Delta G_{45}$ (kcal/mol) |
|---|---|---|---|---|
| *set 1* | AATATTGACAAAATTATATAAAAGTTTTT | 10.0 | 48 | 24.14 |
| *set 2* | AAATATATAACGTATCCTGCGTGTACACAT | 33.3 | 56 | 30.76 |
| *set 3* | CCAAATTGCGCAATTTCGCCACCATGTCAG | 50.0 | 63 | 36.98 |
| *set 4* | TCCATGTCAGCCTTGGTTGACATGTAGCGT | 50.0 | 63 | 36.61 |
| *set 5* | CGCAAAAGTTCAAAAAGCAGCGGCTGCTCA | 50.0 | 63 | 37.88 |
| *set 6* | TCATCATGGCTTGTCCCCCCTTTCACTAA | 50.0 | 63 | 35.71 |
| *set 7* | CGGCCACACCCCCCTGATGGCGTTACCCAT | 66.7 | 70 | 41.07 |
| *set 8* | CCCGGTGGCACTCCGGCCATGGCGACCGGC | 80.0 | 75 | 45.67 |

**Equation 1:** Equation of the Langmuir model.

$$I = I_0 + \frac{Ace^{\Delta G/RT}}{1 + ce^{\Delta G/RT}} \approx I_0 + Ace^{\Delta G/RT}$$

**Figure 1:** Oligo set 1 (GC content = 10%, Tm = 47°C) as an example for the eight oligo sets, containing 12 oligos with none, one, two, or three mismatches at well-chosen positions. The mismatches are highlighted using a black background. The position of the 12-mer universal spacer is indicated by "TAG". The up most sequence is the fluorescently labeled 70-mer target sequence. Apart from differences in the sequences all the other oligo sets have mismatches positioned as in the example given here.

```
5-AGCTTTCATTTACCCTAATTAAAAACTTTTTATATAATTTTGTCAATATTTAAACATAATCTTGCATATT-3
1#                  3-TTTTTGAAAAATATATTAAAACAGTTATAA-5-TAG-C6amino------
2#                  3-TTTTTGAAATATATATTAAATCAGTTATAA-5-TAG-C6amino------
3#                  3-TTTTTGAAAAATATAATAAAACAGTTATAA-5-TAG-C6amino------
4#                  3-TTTTTGTAAAATATATTAAAACAGTTATAA-5-TAG-C6amino------
5#                  3-TTTTTGAAAAATATATTAAAACACTTATAA-5-TAG-C6amino------
6#                  3-TTTTTGTAAAATATATTAAAACACTTATAA-5-TAG-C6amino------
7#                  3-TTTTTGTAAAATATTTTAAAACAGTTATAA-5-TAG-C6amino------
8#                  3-TTTTTGAAAAATATAATAAAACACTTATAA-5-TAG-C6amino------
9#                  3-TTTTTGTAAAATATAATAAAACACTTATAA-5-TAG-C6amino------
10#                 3-TTAATTGTAAAATATATTAAAACAGTTATAA-5-TAG-C6amino------
11#                 3-TTTTTGAAAAATTTATAAAAACAGTTATAA-5-TAG-C6amino------
12#                 3-TTTTTGAAAAATATATTAAAACACTTAAAA-5-TAG-C6amino------
```

**Figure 2:** Reproducibility plots of background-corrected log intensities of replicate hybridizations. The data points in the plots represent the average value for all twelve spots per oligo. Panel A: using all eight targets at a concentration of 3000 pM; panel B: only using target 5 (in black), concentration = 380 pM; panel C: only using target 3 (in black), concentration = 3.8 pM.

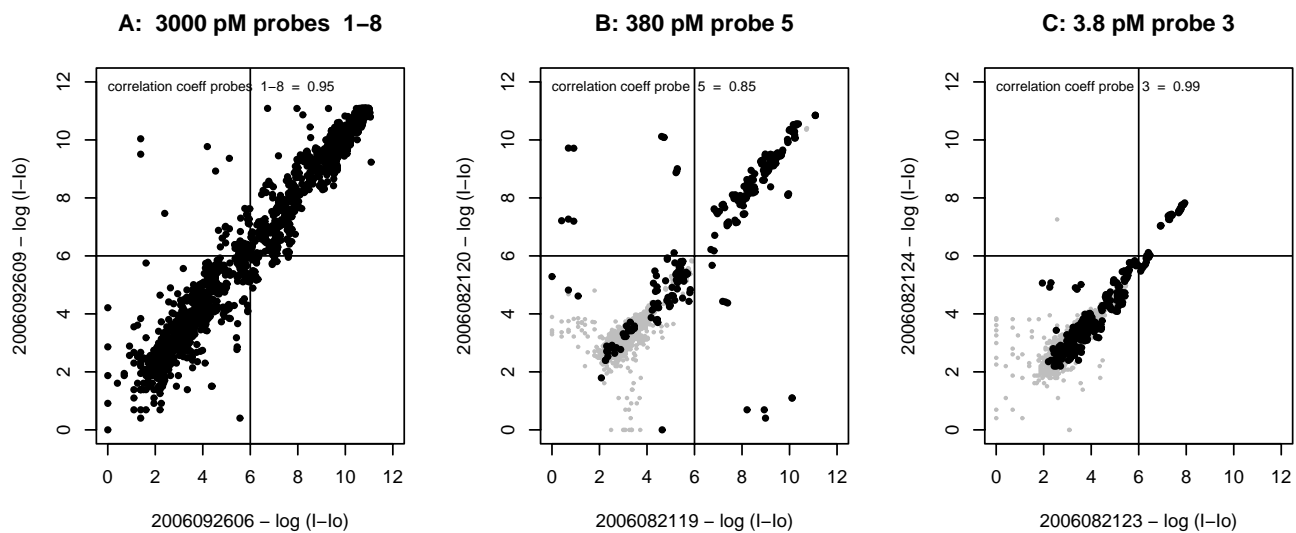

**Figure 3:** Bar charts showing log intensities for six types of oligos. Each bar chart shows the log intensity of one type of oligo (30-mer and 42-mer) in all eight oligo sets. The average of 12 spots gives the height of the bars, the flags indicate a 95% confidence interval. Light grey bars denote intensities that are below the background cut-off, defined as the average background signal for all 12 spots plus twice the standard deviation of this average background signal. The name of the oligos is composed of three numbers, separated by an underscore: the first points to the oligo set number, the second is the number of the oligo in each set (see also Figure 1 and supplementary data) and the third gives the length of the oligo: oligo length 42 stands for the 30-mer oligo with 12-mer spacer attached.

Fig. 3A: oligo 1, PM; Fig. 3B: oligo 3, 1 MM in the middle; Fig. 3C: oligo 10, 2 MM and perfect matching stretch of 23bp; Fig. 3D: oligo 6, 2 MM with a perfect matching stretch of 16 bp in between; Fig. 3E: oligo 11, 2 MM in the middle; Fig. 3F: oligo 9, 3 MM.

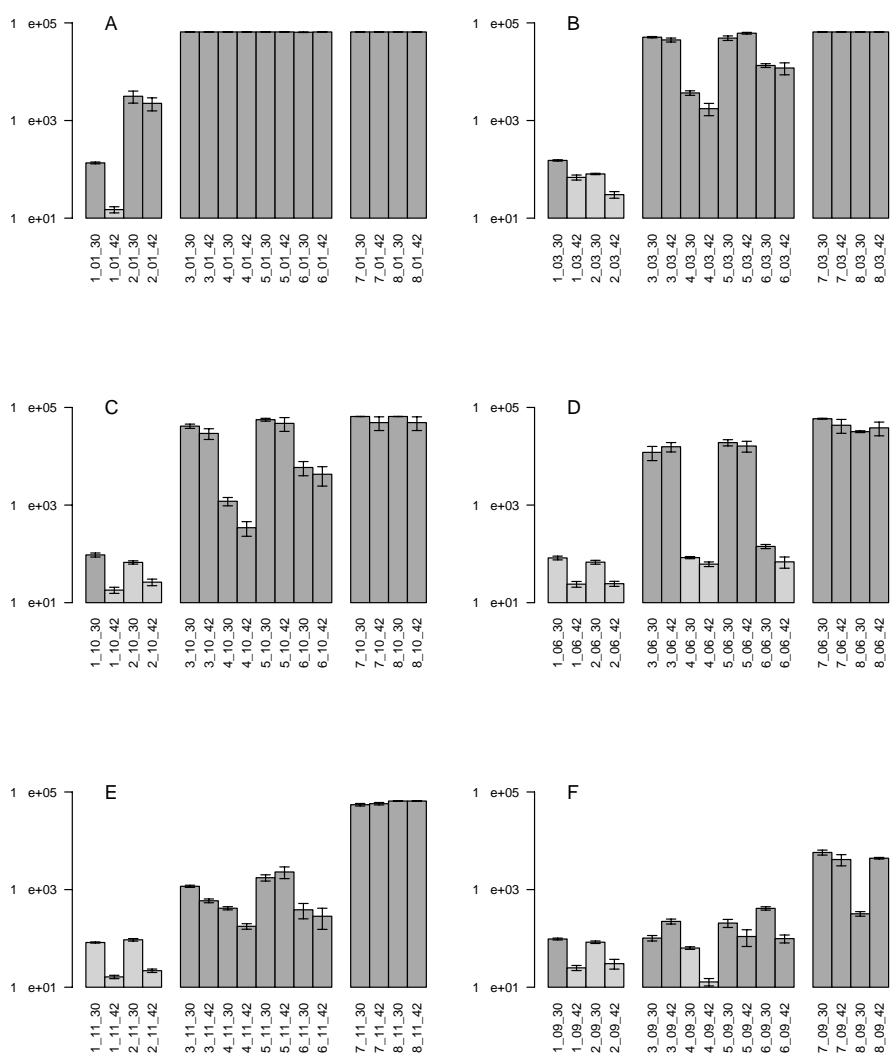

**Figure 4:** Two graphs (a and b) representing log-scale background-corrected hybridization intensities in function of the binding free energy (ΔG) for two hybridizations where all eight targets were used at a concentration of 3000 pM. Only the data points for the oligo sets with a GC content of 50% are plotted. The data points in each graph represent the average of 12 replicate spots of the oligo. The six different kinds of oligos regarding the amount of mismatches and their relative positions, are indicated by the six symbols as represented in the legend, where PM stands for the perfect matching oligo 1, 1MM for oligos 3, 4 and 5 containing 1 mismatch, 2MM-15 for oligos with two mismatches with an inter-mismatch distance smaller than 15 (oligos 2 and 11), 2MM15 for oligos 6, 7 and 8 with an inter-

mismatch distance of 15 bases, 2MM23 for oligos 10 and 12 with an inter-mismatch distance of 23 bases and 3MM for oligo 9 containing three mismatches. The solid line has a slope equal to 1/RT, where T is the experimental temperature (T = 45°C = 318K). The horizontal dashed line corresponds to the saturation level.

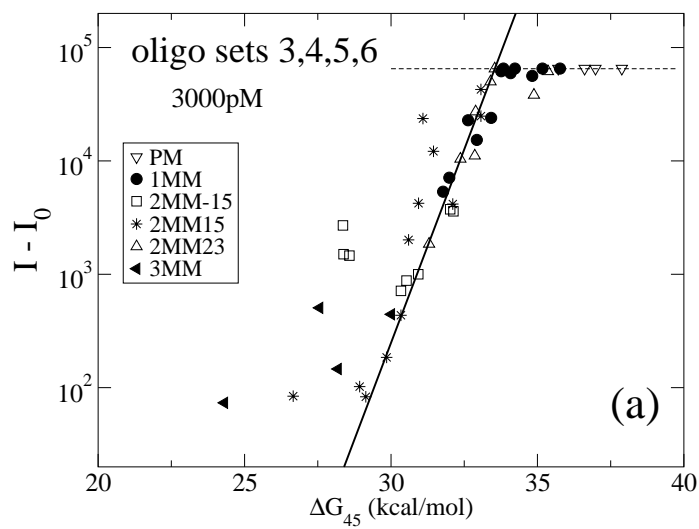

(a)

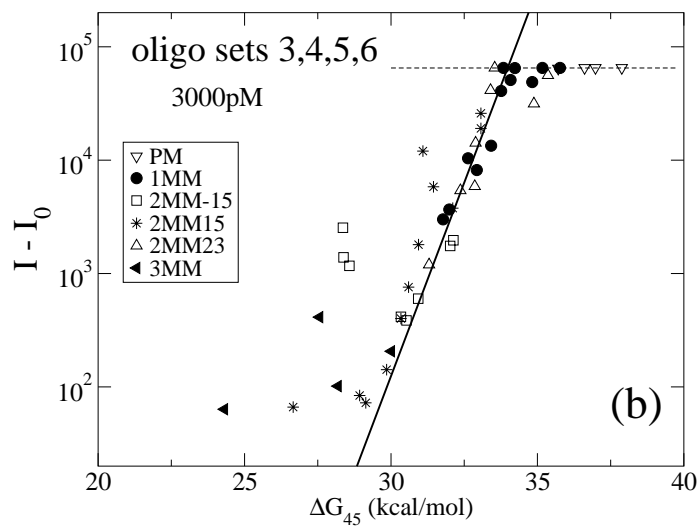

(b)

**Figure 5:** Graph representing log-scale background-corrected hybridization intensities in function of the binding free energy (ΔG) for four hybridizations with a target concentration of 380 pM. Each of the four panels of the Figure represents data from separate hybridizations, each for one of the four oligo sets with a GC content of 50%. The data points in each graph represent the average of 12 replicate spots of the oligo. The six different kinds of oligos regarding the amount of mismatches and their relative positions, are indicated by the six symbols as represented in the legend, where PM stands for the perfect matching oligo 1, 1MM for oligos 3, 4 and 5 containing 1 mismatch, 2MM-15 for oligos with two mismatches with an inter-mismatch distance smaller than 15 (oligos 2 and 11), 2MM15 for oligos 6, 7 and 8 with an inter-mismatch distance of 15 bases, 2MM23 for oligos 10 and 12 with an inter-mismatch distance of 23 bases and 3MM for oligo 9 containing three mismatches. The solid line has a slope of 1/RT, the horizontal dashed line corresponds to the saturation level.

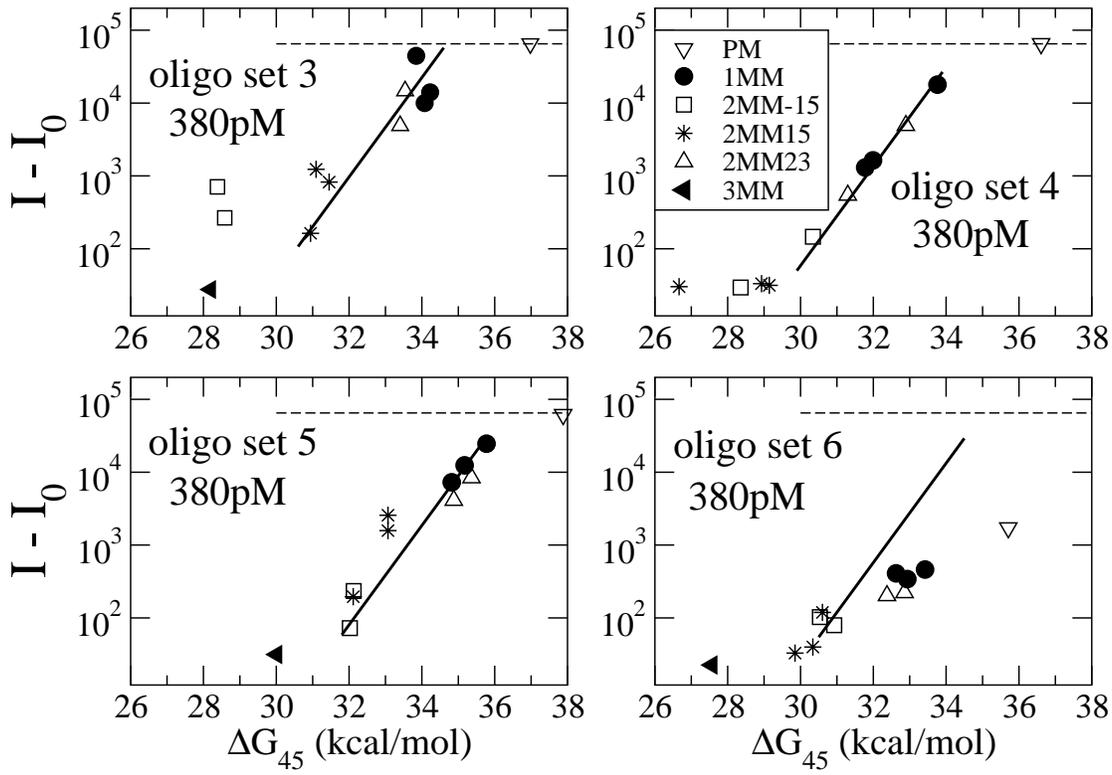

**Figure 6:** Graph representing log-scale background-corrected hybridization intensities in function of the binding free energy (ΔG) for four hybridizations with a target concentration of 3.8 pM. Each of the four panels of the Figure represents data from separate hybridizations, each for one of the four oligo sets with a GC content of 50%. The data points in each graph represent the average of 12 replicate spots of the oligo. The six different kinds of oligos regarding the amount of mismatches and their relative positions, are indicated by the six symbols as represented in the legend, where PM stands for the perfect matching oligo 1, 1MM for oligos 3, 4 and 5 containing 1 mismatch, 2MM-15 for oligos with two mismatches with an inter-mismatch distance smaller than 15 (oligos 2 and 11), 2MM15 for oligos 6, 7 and 8 with an inter-mismatch distance of 15 bases, 2MM23 for oligos 10 and 12 with an inter-mismatch distance of 23 bases and 3MM for oligo 9 containing three mismatches. The solid line has a slope of 1/RT, the horizontal dashed line corresponds to the saturation level.

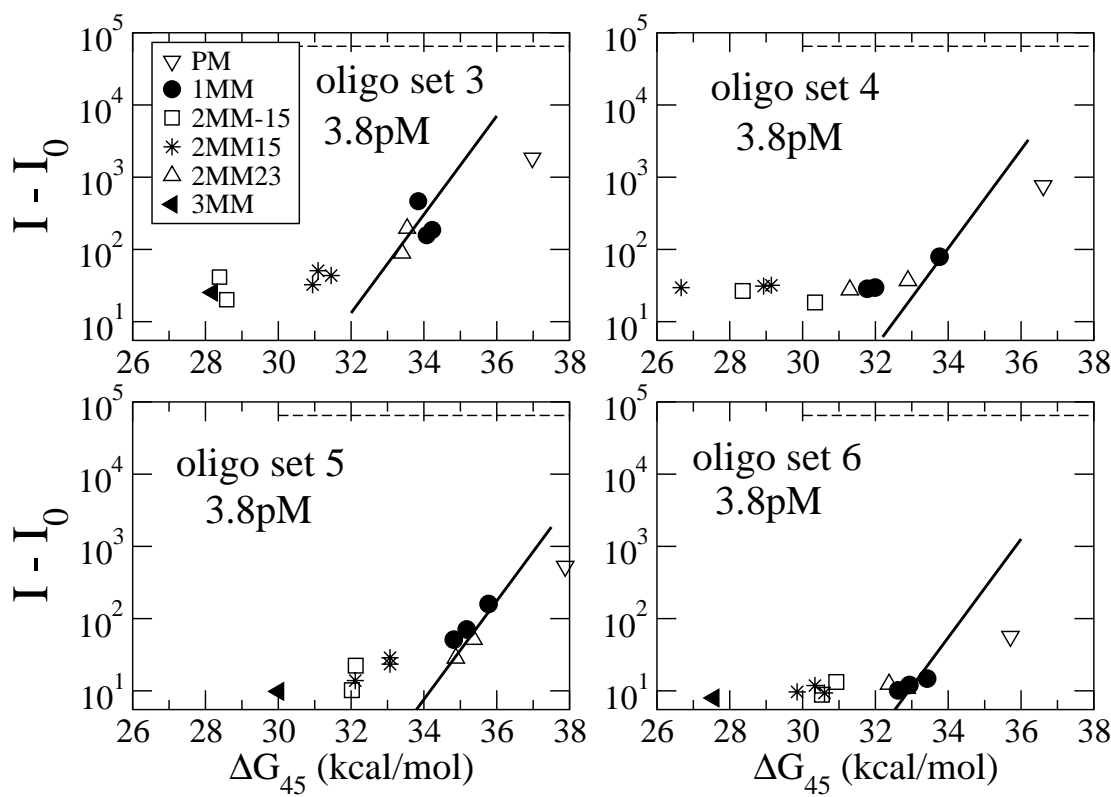